# Taking a Stand or Standing Back: Brand Response Strategy during Social Activism

**Jingjing Li** (Contact Person)

Andersen Alumni Associate Professor of Commerce
McIntire School of Commerce
University of Virginia
jl9rf@virginia.edu

**Nicole Votolato Montgomery**

Professor of Marketing
McIntire School of Commerce
University of Virginia
nvmontgomery@virginia.edu

**Reza Mousavi**

Assistant Professor of IT & Innovation
McIntire School of Commerce
University of Virginia
mousavi@virginia.edu

**Abstract**

With the proliferation of social activism online, brands face heightened pressure from consumers to publicly address these issues. Yet, the optimal brand response strategy (i.e., whether and how to respond) in these contexts remains unclear. This research investigates consumers' reactions to brand response strategies (e.g., engage vs. not) during social activism and offers potentially effective responses that brands can employ to engage in these issues. By analyzing real-world data collected from Twitter and conducting four randomized experiments, this research discovers that brand relationship type (exchange, communal) affects consumers' brand evaluations in the wake of social activism. Communal (vs. exchange) brands are evaluated less favorably when they do not respond or utilize a low-empathy response. This difference is attenuated when brands employ a high-empathy response. These findings are attributable to consumers' perceptions of whether the brand's response strategy complies with relationship norms during social activism. The effects persist across activism events that vary in their political polarization. This research contributes to the literatures on brand engagement in social activism, brand relationships, and crisis communication. The findings also offer guidance to practitioners on crafting response strategies during social activism and aid activists in securing brand support for societal benefits.

*Keywords*: brand engagement in social activism, brand relationship norms, crisis communication, firm-generated content, #MeToo, randomized experiments

**Introduction**

Brands are facing heightened pressure from consumers to engage in social activism (Moorman 2020; Sarkar and Kotler 2018; Vredenburg et al. 2020). *Social activism* is collective action among individuals, groups, and organizations to bring about social, policy, and legislative changes related to a social issue (e.g., MeToo, Black Lives Matter; Tilly and Wood 2020). Given social media's crucial role in disseminating information during these movements (Valenzuela 2013), consumers are now exposed to social activism on an unprecedented scale (Tang and Lee 2013). This heightened exposure also underscores the value of brands' communication channels in expanding activism reach and the role of brand-activist collaborations in enriching operational practices. As such, an estimated 60 to 70% of consumers believe that brands need to respond to social issues, and most prefer such responses to appear on social media (Sprout Social 2019).

Brands' engagement on social issues can impact consumers' brand evaluations and purchase decisions (Hou and Poliquin 2023; Moorman 2020; Neureiter and Bhattacharya 2021; Vredenburg et al. 2020). In extreme cases, consumers may launch buycotts or boycotts to vigorously respond to a brand's stance on an issue, even when brands choose not to declare their position on an issue publicly (Fernandes 2020; Jungblut and Johnen 2021; Neilson 2010). These factors have led an increasing number of brands to consider engaging in conversations about social issues, particularly on social media channels. As the chief executive officer (CEO) of Edelman commented, "[b]rands are now being pushed to go beyond their classic business interests to become advocates" (Edelman 2018).

Recent work from marketing and other disciplines has started to examine whether engaging in social activism is advantageous for brands, but the findings are inconsistent. See Web Appendix A for a review. Some studies have shown that brand engagement in social issues negatively impacts social media engagement (Wang et al. 2022), brand attitudes (Mukherjee and Althuizen 2020), brand choices (Hydock, Paharia, and Blair 2020), and stock performance (Bhagwat et al. 2020). Bhagwat et al. (2020) suggest that this occurs because stakeholders believe that such engagement implies that brands are diverting resources away from profit-generating activities. In contrast, other studies have shown that engagement in social issues benefits brands by enhancing their influence on public policy-making (Werner 2015), increasing purchases (Buell and Kalkanci 2021; Chatterji and Toffel 2019), and reducing activist challenges and boycotts (McDonnell 2016). Still, other work has documented brand differences in social activism engagement (Hydock, Paharia, and Blair 2020; Wang et al. 2022). For instance, Wang et al. (2022) found that the negative impact of brand engagement in social issues could be attenuated for brands with a history of posting prosocial content on social media platforms.

Even when brands choose to engage in social activism, there is prevailing uncertainty about the effective response strategies they should employ. See Web Appendix A. While some studies have emphasized the need to underscore tangible actions and identify specific response sources (like the CEO) in brand responses (Chatterji and Toffel 2019; Vredenburg et al. 2020), others have suggested that consumers may react negatively to such responses (Bhagwat et al. 2020). Real-world instances have also yielded inconsistent consumer reactions. For instance, Nike was praised by many consumers for its "[L]et's be part of the change" tweet addressing racial equity (Netimperative 2020), but Pepsi encountered severe backlash for its response to the same issue, as it was viewed as giving lip service to improving product sales (Victor 2017). As a result, despite the heightened consumer expectations for brand engagement in social activism (Lin 2018), brands often hesitate to engage (Moorman 2018, 2020).

Against this backdrop of diverse findings, scholars have recently called for more nuanced theories and concrete empirical studies to better understand consumers' reactions to brand engagement in social issues (Moorman 2020; Neureiter and Bhattacharya 2021; Vredenburg et al. 2020). We respond to this call by integrating research on brand engagement in social issues, brand relationships, and crisis communication. Specifically, we clarify the brand response strategy (e.g., engage in social issues vs. not) by investigating the impact of such strategies on consumers' brand evaluations. We also elucidate the specific responses most effective for brand engagement in social activism. We utilize brand evaluations as our focal dependent measure to mirror the prior literature on brand relationship norms and brand engagement in social issues, which focuses on consumers' reactions to brands (e.g., Aggarwal 2004; Goh, Heng, and Lin 2013; Mukherjee and Althuizen 2020). This variable can also drive other consequential outcomes, such as purchase intent, word-of-mouth (WOM), and social media engagement (Ajzen 1991; Ajzen et al. 2018; Glasman and Albarracín 2006). We include them as additional dependent measures in one of our studies.

We theorize that the effect of a brand's response strategy on evaluations of the brand in the wake of social activism is related to consumers' relationship with the brand–i.e., exchange or communal–and the norms that guide it (Aggarwal, 2004; Clark and Mills, 1993). Exchange relationship norms create expectations that brands will effectively provide the transactional outcomes (e.g., quality product or service) for which consumers are paying. On the other hand, communal relationship norms create expectations that a brand will be helpful, supportive of others, or attentive to others' feelings (Aggarwal 2014, p. 50). Building on this prior work, we argue that consumers have higher (lower) expectations of a communal (exchange) brand's social responsibility and thus engagement in social issues (Bolton and Mattila 2015; Du, Bhattacharya, and Sen 2011; Jin and Lee 2013). When a brand's response strategy violates the norms and expectations of the relationships, consumers evaluate the brand less favorably compared to when there are no violations (Aggarwal 2004; Aggarwal and Larrick 2012; Aggarwal and Law 2005).

Across a secondary data analysis and four controlled experiments, we test our theorizing via two of the most influential Twitter-native social movements in the past decade: #MeToo and #BlackLivesMatter (hereafter referred to as #BLM). We find that consumers' brand evaluations in the wake of social activism vary as a function of brand relationship type (exchange vs. communal brand). We also examine the responses that brands may utilize to engage in social activism and show that the differences in consumer evaluations by brand relationship type can be attenuated with a high empathy response.

Our research contributes to the literature on brand engagement in social activism by clarifying how brand response strategy (i.e., specific response or lack thereof) in the wake of social activism impacts consumers' brand evaluations. The consumer-brand relationship theoretical lens allows us to more carefully unpack the causal chain connecting social activism and brand evaluations and, in doing so, account for apparent inconsistencies in this domain (e.g., Bhagwat et al. 2020; Buell and Kalkanci 2021; Moorman 2020; Wang et al. 2022). Furthermore, our research extends brand relationship norm theory by expanding its boundary conditions. Existing research mainly explores how brand-controlled actions (e.g., customer service issues, product failures) violate or conform to salient relationship norms (Aggarwal 2004; Aggarwal and Larrick 2012; Aggarwal and Law 2005). Our studies suggest that external factors, like social activism, can also impact perceptions of brands' norm compliance, thereby influencing consumers' evaluations. Finally, our work offers marketing professionals guidance on social activism response strategies. We also aid social activists in garnering brand support, thereby enhancing societal influence and the benefits of their spearheaded initiatives.

## Theoretical Development

*Brand Relationship Norms and Social Activism*

We propose that consumers' a priori relationships with brands are critical in understanding consumers' evaluations of these brands in the wake of social activism, thereby furthering our understanding of brand engagement in social activism. Brand relationship norm theory suggests that consumers form different relationships with some brands than others. The difference is guided by the norms governing the giving and receiving of benefits between consumers and brands (Clark and Mills 2012, Clark and Mills 1993). Specifically, consumer-brand relationships may be guided by exchange norms, communal norms, or a combination of both (Aggarwal and Law 2005). Exchange relationship norms emphasize the quid pro quo nature of brand relationships and, consequently, lead consumers to expect a brand to deliver on certain performance outcomes (Clark and Mills 2012). Such norms align with the kinds of interactions that consumers tend to have with brands – i.e., providing payment in exchange for a product or service – and they, therefore, tend to comprise most consumer-brand relationships (Ahluwalia, Burnkrant, and Unnava 2000; Brown and Dacin 1997; Gürhan-Canli and Batra 2004).

Yet, consumers may also attribute human-like qualities to certain brands (Fournier 1998), increasing communal norms' relative influence. Communal norms are more akin to those that guide interpersonal relationships (e.g., friends; Kanouse and Hanson 1972; Wojciszke, Brycz, and Borkenau 1993) and often lead consumers to have similar expectations of brands as they would for individuals – i.e., exhibiting a greater concern for others (Aggarwal 2004). As such, they often expect the brand to adhere to a moral code and show caring and support for others in society (Aggarwal 2004; Kanouse and Hanson 1972; Wojciszke, Brycz, and Borkenau 1993). In contrast to exchange norms, communal norms can differ significantly across brands (Montgomery and Cowen 2020). We refer to brands for which consumer relationships are guided relatively more (less) by communal than exchange norms as *communal* (*exchange*) *brands*. Considerable research has documented benefits for communal (vs. exchange) brands. Communal brands tend to have stronger consumer relationships (Aaker 1997; Aaker et al. 2004; Fournier 1998), more favorable evaluations, greater purchase loyalty, continued consumer support amidst negative publicity (Aaker 1997; Ahluwalia et al. 2000), and more brand equity (Keller 1993). Therefore, many brands strive to build strong communal relationships with consumers

Consumers' evaluations of a brand are based on whether the brand's actions conform to the norms of the relationship (Aggarwal 2004; Aggarwal and Law 2005). Notably, the brand actions studied in the existing literature are often endogenous and directly tied to a brand's products and services, such as charging a fee for customer service (Aggarwal 2004). Consumers react less favorably to a brand when its actions violate the guiding norms of the relationship (Aggarwal 2004). Accordingly, evidence suggests that consumers evaluate communal brands' actions similarly to how they would evaluate those of a person (Kanouse and Hanson 1972; Wojciszke, Brycz, and Borkenau 1993). They are less forgiving of behaviors that appear immoral or show a lack of concern for others because these actions violate expectations that the brand will behave ethically (Montgomery and Cowen 2020). Yet, exchange brands may not be penalized for behaviors that fail to show caring or support for others, since these actions do not violate expectations of performance (Ahluwalia, Burnkrant, and Unnava 2000; Brown and Dacin 1997; Gürhan-Canli and Batra 2004).

We propose that differences in the relative influence of norms for communal versus exchange brands also affect consumer evaluations of brands in the wake of social activism. As discussed, consumers often expect communal brands to show caring and support for others in society, more so than exchange brands (Aggarwal 2004; Kanouse and Hanson 1972; Wojciszke, Brycz, and Borkenau 1993). Social activism often highlights inequalities or discrimination against certain members of society. For example, the #MeToo movement brought attention to sexual harassment victims. Brand engagement in social activism is considered an extension of Corporate Social Responsibility (CSR) (Sarkar and Kotler 2018; Vredenburg et al. 2020). In the wake of social activism, which often centers on supporting disadvantaged groups and advocating for change (Bolton and Mattila 2015; Jin and Lee 2013), brand silence can be interpreted as indifference to injustice and a lack of caring for others, which is inconsistent with communal norms. Thus, when a *communal* brand does not respond to social activism, we expect that consumers will perceive that the brand has failed to comply with relationship norms and, consequently, will evaluate the brand less favorably than in contexts in which there is no social activism. In contrast, communal norms have relatively less influence on consumers' relationships with exchange brands, so social activism should be less important to consumer evaluations. Hence, when an *exchange* brand does not respond to social activism, no norms are perceived to be violated. Therefore, we expect that *exchange* brands will be evaluated similarly when social activism occurs versus when it does not. Consequently, communal brands are expected to be evaluated less favorably than exchange brands in the wake of social activism. Further, consumers' perceptions of a brand's compliance with relationship norms (i.e., norm compliance) should mediate this effect. Therefore, we hypothesize:

H1: When brands do not respond to social activism, consumers will evaluate communal brands less favorably than exchange brands.

H2: Norm compliance will mediate the effects proposed in H1.

***Empathy and Brand Response Strategy***

We contend that if communal brands remain silent in the wake of social activism, they will be evaluated less favorably than exchange brands because they are perceived to violate relationship norms (e.g., caring for others) to a greater extent. A question that remains is how these brands should respond to social activism if they do engage in these social issues. For communal brands, we propose that responses to social activism that demonstrate concern and care for victims of injustice may increase perceived compliance with relationship norms and, in turn, brand evaluations. As such, we explore the effectiveness of responses that vary in empathy.

Empathy is the capacity to recognize, feel, and/or react compassionately to others, which is consistent with communal norms that emphasize care for others (Batson, Fultz, and Schoenrade 1987; Coke, Batson, and McDavis 1978; Ekman 2004; Fournier and Alvarez 2012; Powell and Roberts 2017). Empathy is a multidimensional construct ranging from the lowest level of empathy (i.e., cognitive empathy) to the highest level (i.e., compassionate empathy) (Ekman 2004). Cognitive empathy acknowledges others' issues, whereas compassionate empathy *acts* to help address those issues. As the level of empathy in responses may vary, it is unclear how much empathy brands should utilize when engaging in social activism. Research on crisis communication offers insight into this issue, suggesting that an effective response strategy *matches* evaluators' situational attribution of a crisis (Bundy et al. 2017; Bundy and Pfarrer 2015; Coombs and Holladay 2002, 2010). An organization perceived as culpable should employ an *accommodative* response that acknowledges the firm's role in a crisis and seeks to address the interests of affected parties (Bundy and Pfarrer 2015; Cowen and Montgomery 2020).

According to our theorizing, consumers are more likely to expect communal brands to engage in social activism than an exchange brand – i.e., they attribute more responsibility to the former in these contexts. Hence, for communal brands, a high empathy response (e.g., acting to address social issues) should result in more favorable brand evaluations than a low empathy response (e.g., merely acknowledging social issues) because it conforms to communal relationship norms and matches the perceived degree of culpability. In contrast, for exchange brands, response empathy should be less important to brand evaluations, given the lesser influence of communal norms. Therefore, the positive effect of a high empathy response to social activism, relative to a low empathy response, on consumer evaluations will be greater for communal brands (vs. exchange brands). If our theorizing is correct, a low empathy response should result in less favorable evaluations of communal (vs. exchange) brands, whereas a high empathy response should attenuate this difference. Thus, we hypothesize:

H3: When brands respond to social activism on social media, response empathy will moderate the effect proposed in H1, such that H1 will be supported with a low empathy response but attenuated with a high empathy response.

## Roadmap of Studies

To test our hypotheses, we selected #MeToo as the primary and #BLM as the secondary social activism context in our investigation for two reasons. First, #MeToo and #BLM are both prevalent activism events underscoring social justice violations. However, they diverge greatly in terms of their levels of political polarization (i.e., how politically divisive the issues are; Pew Research 2019). This divergence allows us to investigate the persistence of our predicted effects across events with varying degrees of political polarization. Second, unlike #BLM, most brands did not respond to #MeToo on Twitter, providing a relatively homogeneous response scenario for analysis. See Web Appendix D. This enables us to use the Twitter data in Study 1 to examine the impact of the lack of response to #MeToo on consumer brand evaluations. We chose Twitter as our focal social media platform due to its acclaimed position as a top platform for social activism (Jackson, Bailey, and Welles 2020) and its widespread use in prior research on social movements (Li et al. 2021).

Using secondary data analysis and randomized experiments, we conducted five studies to test our hypotheses. The secondary data analysis (Study 1) serves as an exploratory analysis to establish an association between the activism context and brand evaluations for #MeToo (H1). To address possible identification and endogeneity issues, such as a lack of control for the heterogeneity in brands, consumers, and responses in this study, we also use four randomized experiments (Studies 2-5). The experiments also reveal the underlying causal mechanism—brand relationship norm compliance. Specifically, Study 2 documents the effect of brand relationship norms on consumers' brand evaluations in the wake of #MeToo social activism (H1) by manipulating communal and exchange norms for a fictitious brand, mirroring prior work on brand relationship norms (Aggarwal 2014). Study 3 extends the observed effects to real communal and exchange brands with existing consumer relationships (H1) and further supports our theorizing by demonstrating the moderating role of the activism context (i.e., social activism vs. no activism). Study 4 explores the robustness of our findings by investigating two different social activism events (#MeToo and #BLM) that vary in their political polarization. Studies 2-4 also offer support for the proposed causal mechanism (i.e., norm compliance; H2). Study 5 examines the effectiveness of different response strategies that vary in their level of empathy (H3) and further confirms our proposed underlying mechanism (H2).

**Study 1**

Study 1 aims to explore whether consumer evaluations of communal versus exchange brands change in the wake of a social activism event using real data collected from Twitter. To this end, we conducted a fixed-effects regression on a monthly panel dataset comprised of different brands (communal vs. exchange brands) and brand evaluations (monthly sentiment of the tweets mentioning these brands). Of note, the goal of this study is not to provide causal inferences, but rather to document the external validity of our predicted effects and motivate the design of the subsequent randomized experiments.

*Data and Model*

We used the *Twitter API for Academic Research* to collect tweets mentioning the official Twitter handles of 118 popular Business-to-Consumer (B2C) brands from 10/01/2015 to 10/2019. This period covers two years before and two years after the first #MeToo tweet posted by Alyssa Milano in October 2017. Milano's tweet came in the wake of a New York Times article featuring Ashley Judd's allegations against Harvey Weinstein. This tweet resulted in 5.8 billion reactions (likes, retweets, etc.) within the first few days. The brands were selected based on the survey data from Xu et al. (2016), which included consumer responses to items related to communal and exchange relationship norms. Due to the API's rate limit, we randomly sampled 400 tweets for each brand per week. All tweets (2,137,884 tweets) were collected in May and June 2021.

To verify that consumers expect brands to engage in social activism, we first conducted a text analysis on the tweets posted by users (other than brands) that mentioned brands and #MeToo. We found that 104 out of 118 brands were mentioned in tweets related to #MeToo. See Web Appendix D for a detailed synthesis of these tweets. However, after examining the tweets posted by brands, we found only two brands directly responded to #MeToo within our sampling period. See Web Appendix D. As a result, this data represents a scenario in which the majority of brands did not actively respond to social activism, providing us an opportunity to use secondary data to investigate the impact of a homogenous response strategy (i.e., no response) to social activism on consumers' brand evaluations.

We measured brand evaluations by the sentiment of tweets that mentioned these brands (*Brand_Sentiment* hereafter). We used the Python package Stanza, which uses advanced deep learning models such as convolutional neural networks to classify a tweet into positive (a value of 1) or negative (-1). We then aggregated sentiment scores to the brand-month level by taking the mean, resulting in a monthly panel dataset of brands from 10/2015 to 10/2019.

Based on the items from Aggarwal and Larrick (2012) and Xu et al. (2016), we generated four measures for determining the brand relationship types (communal vs. exchange brand): a communal score (used as the dependent variable in Model 1- Table 1), a binary communal score via median split (Model 2- Table 1), a communal score subtracted by an exchange score (Mode3 2- Table 1), and a binary communal-minus-exchange score via median split (Model 4- Table 1). See Web Appendix D for details about these measurements.

We then ran the following fixed effects regression model:

$$Brand\_Sentiment_{i,t} = \alpha_i + \gamma_t + \beta \times MeToo_t \times Brand\_Relationship_i + \varepsilon_{i,t} \qquad (1)$$

where $i$ is the indicator for a brand, $t$ is the indicator for time (month), $\alpha_i$ is the brand fixed-effects, $\gamma_t$ is the time-fixed effects, $\beta$ is the coefficient of interest, *MeToo* is a dummy variable that is zero before October 2017, *Brand_Relationship* measures the salience of relationship norms using one of the four methods described earlier, either as a continuous score for communal/exchange norms or as a binary variable for communal/exchange brands, and *Brand_Sentiment* is the average monthly sentiment score of tweets that mentioned a brand (excluding brands' self-mentions).

***Findings***

Figure 1 visualizes the trends of average *Brand_Sentiment* for communal versus exchange brands before and after the first #MeToo tweet. Communal brands had a higher sentiment score before #MeToo, which is consistent with the literature that consumers often evaluate communal brands more favorably than exchange brands (Aaker 1997, Aaker et al. 2004, Fournier 1998, Keller 1993). However, this gap was closed after #MeToo, primarily driven by a more significant drop in sentiment for communal brands than exchange brands.

Figure 1 Sentiment Trends for Communal (Red) and Exchange Brands (Blue) Two Years Before and Two Years After the First #MeToo Tweet.

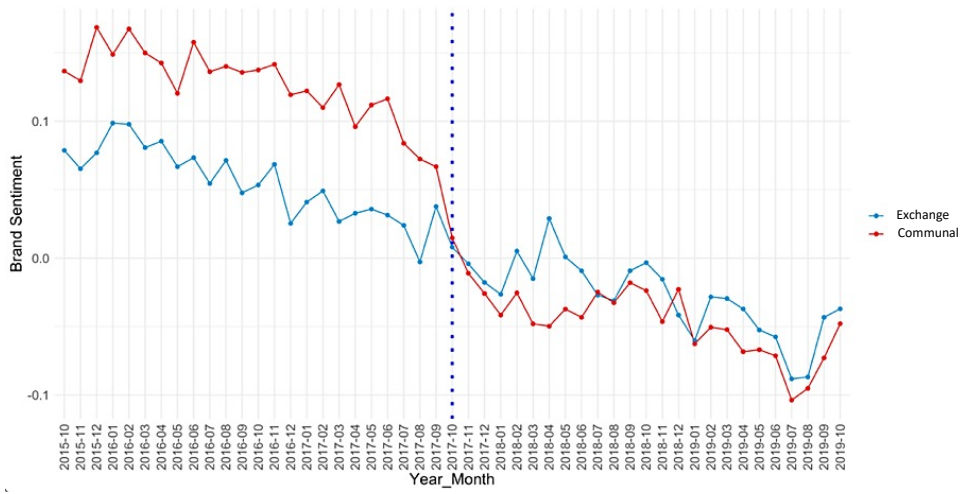

Table 1 Results of the Fixed Effects Model.

|  | **Model 1** | **Model 2** | **Model 3** | **Model 4** |
|---|---|---|---|---|
| *MeToo × Brand_Relationship* | -0.086*** (0.020) | -0.112*** (0.029) | -0.080*** (0.013) | -0.111*** (0.029) |
| Time-fixed effects | ✓ | ✓ | ✓ | ✓ |
| Brand-fixed effects | ✓ | ✓ | ✓ | ✓ |
| Robust | ✓ | ✓ | ✓ | ✓ |
| F-Stat. | 33.987 P < 0.001 | 34.142 p < 0.001 | 39.093 p < 0.001 | 34.077 p < 0.001 |
| R-Sq. | 0.230 | 0.231 | 0.255 | 0.230 |

*Notes.* Model 1 uses a continuous communal score as the brand relationship type. Model 2 uses a binary communal score via median split. Model 3 uses a continuous communal-exchange score. Model 4 uses a binary communal-exchange score via median split. *** significant at 0.001. Robust standard errors clustered at the brand level are in parentheses.

Table 1 provides statistical evidence. The coefficient $\beta$ is negative and significant in all four models, suggesting that communal brands received more negative brand sentiment than exchange brands after #MeToo. These results are aligned with our first hypothesis (H1). However, as discussed before, we acknowledge that this analysis suffers from potential identification and endogeneity issues due to its descriptive nature, and therefore, we conducted Studies 2-4 to further test our first hypothesis.

**Study 2**

Study 2 had two primary objectives. First, we wanted to see whether, consistent with our predictions in H1, individuals would penalize a communal brand (i.e., less favorable brand evaluation) for not responding to social activism (e.g., #MeToo), but not an exchange brand. Second, we wanted to examine the role of norm compliance in mediating the effect of brand relationship on consumers' brand evaluation in the context of social activism, as predicted in H2. All studies were approved by the Institutional Review Board at the authors' university.

**Method**

*Stimulus materials pretest*

Study 2 was designed to parallel research by Aggarwal and colleagues (Aggarwal 2004, Aggarwal and Larrick 2012, Aggarwal and Law 2005, Aggarwal and Zhang 2006), which has demonstrated that consumers' relationships with brands shape their overall evaluations. We therefore chose to use the same communal and exchange relationship scenario descriptions from Aggarwal (2004) to vary participants' perception of their relationship with a fictitious bank (i.e., Grove Bank) to be communal or exchange. This approach allowed us to utilize the same brand consistently across conditions, and, in doing so, ensure that any observed effects are not attributable to other potential differences by brand relationship (e.g., product categories, prices and quality, brands' prior response strategies, and consumers' prior experiences).

To increase our confidence in this approach, we ran a pretest to ensure that these scenarios yielded differences in the perception of the brand relationship, as intended, and not other constructs. Participants in the pretest ($N = 74$, 41.9% female, 56.8% male, 1.4% prefer not to say, $M_{Age} = 39.39$, $SD = 10.46$, Age range = 22-70 years)[1] and main study were recruited from Amazon Mechanical Turk via CloudResearch in exchange for monetary compensation. For Study 2, all respondents were also current bank users, which they indicated via a prescreening question. Each of these studies was completed via an online survey.

Pretest respondents were randomly assigned to read either the communal or exchange relationship scenario description. They then completed the seven-item communal relationship measure and the three-item exchange relationship measure from Aggarwal (2004). We expected that the communal scenario would lead to higher communal (vs. exchange) perceptions than the exchange scenario. To test this, we created a Communality Index by subtracting the average of the exchange measure items from the average of the communal measure items. We further assessed the brand relationship by asking respondents to consider the extent to which Grove Bank would be a more communal person (2 items) or exchange person (2 items) if it were alive (Aggarwal 2004). We created a Communal Person Index by subtracting the average of the exchange items from the average of the communal items. For both the Communality Index and the Communal Person Index, higher (lower) numbers were more indicative of a communal (exchange) brand relationship. As expected, the communal scenario description yielded higher scores than the exchange scenario for both the Communality Index ($M_{Communal}$ = -0.33, $SE$ = 0.16 vs. $M_{Exchange}$ = -0.87, $SE$ = 0.16; $t(72)$ = -2.41, $p$ = .02, $d$ = .56), and the Communal Person Index ($M_{Communal}$ = -0.11, $SE$ = 0.43 vs. $M_{Exchange}$ = -1.32, $SE$ = 0.42; $t(72)$ = -1.99, $p$ = .05, $d$ = .46). To confirm that the relationship scenarios did not also lead to differences on evaluations of the brand or affect, respondents completed a 3-item measure of brand attitudes and the 10-item International Positive and Negative Affect Schedule Short Form (I-PANAS-SF; Thompson 2007). There were no significant differences in the brand attitudes or I-PANAS-SF measures, all $t(72)$ < 1.06, $p$ > .29, $d$ < .25. See Web Appendix B Table W2 for details on the pretest measures.

---

[1] Three respondents in the Study 2 pretest were excluded for completing the survey in less than 80 seconds. Average completion time for remaining respondents was 3.0 minutes ($SE = 9.92$ seconds). The reported sample size reflects the exclusion of these respondents.

We contend that H1 and H2 arise from different expectations of CSR for communal (vs. exchange) brands. This is because brand engagement in social activism is considered an extension of CSR (Sarkar and Kotler 2018; Vredenburg et al. 2020). To confirm this theorizing, we conducted a second pretest to ensure that the CSR expectations indeed differed for the two brands. Pretest respondents ($N$ = 167, 53.3% female, 44.9% male, 1.2% non-binary/third gender, 0.6% prefer not to say, $M_{Age}$ = 43.22, $SD$ = 10.95, Age range = 30-72 years, CloudResearch) were randomly assigned to read either the communal or exchange bank scenario description. They then completed a 3-item measure for expectations of social responsibility (Gürhan-Canli and Batra 2004). As predicted, respondents reported higher expectations of social responsibility at Grove Bank after reading the communal relationship scenario ($M$ = 6.04, $SE$ = 0.13) than the exchange scenario ($M$ = 5.44, $SE$ = 0.13), $t(165)$ = -3.24, $p$ < .01, $d$ = .50.

*Participants, design and procedure*

In Study 2 ($N$ = 264, 57.6% female, 41.7% male, 0.8% prefer not to say, $M_{Age}$ = 43.92, $SD$ = 11.03, Age range = 30-73 years, CloudResearch), we manipulated communal versus exchange brand relationships between subjects.[2] The main study used a procedure similar to the one used in the pretests. Respondents were randomly assigned to read either the communal or exchange brand relationship scenario description. Participants in all conditions then read a scenario and were asked to imagine themselves experiencing what was described. In the scenario, they were told that they were scrolling on Twitter and came across a trending conversation about #MeToo comprised of descriptions of harassment and discrimination towards female employees at different organizations. Finally, they read that Grove Bank posted a generic tweet around the same time as this conversation but did not respond to the #MeToo conversation, after which they were presented with the Grove Bank tweet. See Web Appendix C.

After reading the scenario and the generic tweet, participants reported their brand attitudes (Montgomery and Cowen 2020) and their perceptions of the brand's norm compliance (Coombs and Holladay 2002). Next, they completed a measure of their position on the #MeToo movement (Sen and Bhattacharya 2001) and their political ideology (liberal vs. conservative; Smith et al. 2018). These measures were included as robustness checks in our analyses; we wanted to account for any individual differences related to our focal social activism issue or partisanship. See Web Appendix B Table W3 for details on measures. Finally, participants provided their demographic information before being thanked and debriefed.

*Findings*

An ANOVA revealed the predicted main effect of the brand relationship on brand attitudes, $F(1, 262) = 5.09$, $p = .03$, $\eta_p^2 = .02$. Respondents evaluated the communal brand less favorably ($M = 5.78$, $SE = 0.11$) than the exchange brand ($M = 6.12$, $SE = 0.11$). Overall, these findings support H1.

---

[2] Three respondents in Study 2 were excluded for completing the survey in 100 seconds or less. Average completion time for remaining respondents was 4.4 minutes ($SE = 9.20$ seconds). The reported sample size reflects the exclusion of these respondents.

*Robustness checks: respondent gender, #MeToo position, and political ideology*

Although we did not hypothesize effects by respondent gender, these are important to examine since our focal social activism pertains to gender discrimination and could, consequently, impact our pattern of results. Respondent gender could affect the generalizability of our findings, further underscoring the importance of understanding any effects it may have on our findings. Therefore, we conducted an additional analysis to investigate the role of respondent gender on brand evaluations in the wake of social activism. Specifically, we conducted an ANOVA to examine the interactive effect of brand relationship and respondent gender on brand attitudes. The interaction coefficient was insignificant, $F(2, 258) = 1.41$, $p = .25$, $\eta_p^2 = .01$.

To rule out the possibility that alternative mechanisms proposed in prior studies (e.g., issue position, partisanship) may account for our results, we also conducted analyses to examine the main effect of brand relationship on brand attitudes with the inclusion of covariates (as controls). The main effect of brand relationship on brand attitudes remained significant with #MeToo position included as a control, $F(1, 261) = 3.74$, $p = .05$, $\eta_p^2 = .01$, and with political ideology included as a control, $F(1, 261) = 3.87$, $p = .05$, $\eta_p^2 = .02$. Combined, these suggest the observed effects are not attributable to individual differences in issue or political stance.

*Mediation analysis*

In H2, our contention is that for a communal brand that fails to respond to social activism, individuals' perceptions of the brand's compliance with relationship norms are lower than for an exchange brand that does not respond. This, in turn, results in lower brand evaluations for communal versus exchange brands. To investigate the causal role of norm compliance, we conducted the mediation analyses using PROCESS, with 5000 bootstrap samples (Model 4; Hayes 2017). In our analysis, brand relationship (0 = *exchange brand*, 1 = *communal* *brand*) was the predictor, norm compliance was the mediator, and brand attitudes was the dependent variable. See Web Appendix B Table W4 for norm compliance measure means. As predicted, the indirect path of brand relationship on brand attitudes through norm compliance was negative and significant (*b* = -0.23, *SE* = 0.11, *95% CI* = [-0.46, -0.03]), suggesting a violation of communal norms if communal brands did not respond to social activism. See Figure 2. These findings support the mediating role of norm compliance (H2) on our observed effects.

Figure 2: Norm Compliance of Response Strategy Mediates the Effect of Communal vs. Exchange Brands on Brand Attitudes in Study 2.

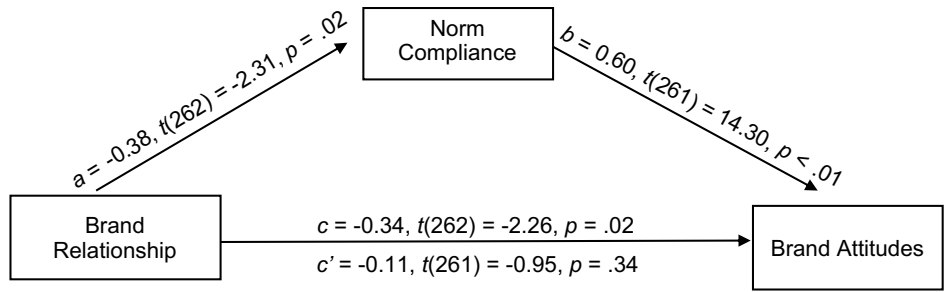

*Note.* N = 264. Indirect mediation effect: *b* = -0.23, *SE* = 0.11, *95% CI* = [-0.46, -0.03].

**Study 3**

The main goal of Study 3 was to replicate the findings from Study 2 (H1 and H2) using actual brands to show that our proposed effects are not limited to hypothetical brand relationship scenarios. To that end, we used real communal and exchange brands. To further increase the external validity of our effects, we presented participants with a series of tweets about #MeToo, rather than describing the #MeToo Twitter conversation. Finally, to provide additional support for our theorizing that consumers penalize communal brands but not exchange brands when brands do not respond to social activism, we also manipulated the context such that participants evaluated the brands in the presence or absence of a social activism context. We expected individuals to respond less favorably to the communal (vs. exchange) brands in a social activism context but not differ in their responses in a context in which there is no social activism.

*Method*

*Stimulus materials pretest*

Study 3 presented information to respondents in the form of a brand tweet from an exchange or communal brand. As such, we ran a pretest to facilitate the selection of the brands to be used in the study. A second goal of the pretest was to verify that the communal (vs. exchange) brand indeed raised expectations of social responsibility as theorized. The pretest used the same respondent sample as the main study (CloudResearch).

In the pretest ($N = 79$, 46.8% female, 53.2% male, $M_{Age} = 39.32$, $SD = 11.33$, Age range = 24-67 years),[3] each respondent was randomly assigned to evaluate seven brands from a list of fourteen different well-known brands. See Web Appendix C Table W11 for brand selection criteria. For each brand, respondents completed the two-item communal relationship measure and the two-item exchange relationship measure from Aggarwal and Larrick (2012). We calculated a Communal Index by subtracting the average of the exchange items from the average of the communal items. Our objective was to choose two brands that differed on the Communality Index – i.e., one with a higher score (communal brand) than the other (exchange brand). We also included measures of brand familiarity (Raju, Unnava, and Montgomery 2009) and brand evaluations (Montgomery and Cowen 2020). We sought to identify two brands that did not differ on these two measures to ensure that any differences we observed in the main study were attributable to differences in the communality of the brand but not individuals' prior experience or a priori perception. See Web Appendix B Table W5 for the pretest measures.

---

[3] One respondent in the pretest was excluded from the analysis for completing the survey in less than 60 seconds, and one respondent was excluded for completing the survey in 14.5 minutes (more than four standard deviations greater than the mean completion time). Average completion time for remaining respondents was approximately 3.5 minutes ($SE = 15.95$ seconds). The reported sample size for the pretest reflects the exclusion of these respondents.

From this pretest, we selected Starbucks as the communal relationship brand and Wendy's as the exchange relationship brand. Our analyses confirmed that these two brands differed on the Communality Index ($M_{Starbucks}$ = -2.21, $SE$ = 0.36 vs. $M_{Wendy's}$ = -3.22, $SE$ = 0.38), $t(79)$ = -1.93, $p$ = .058, $d$ = .43. None of the other measures yielded significant differences between these two brands, all $t$s < 1. The pretest also included a single-item measure (Gürhan-Canli and Batra 2004) of expectations of social responsibility (i.e., *I expect [brand] to be socially responsible*) to ensure that expectations did differ for the two brands, confirming our theorizing. As predicted, respondents reported higher expectations of social responsibility for Starbucks ($M$ = 5.12, $SE$ = 0.27) than for Wendy's ($M$ = 4.08, $SE$ = 0.27), $t(79)$ = -2.70, $p$ = .01, $d$ = .61.

*Participants, design and procedure*

Participants ($N$ = 379, 44.6% female, 54.6% male, 0.8% prefer not to say, $M_{Age}$ = 40.75, $SD$ = 12.36, Age range = 21-78 years) were U.S.-based respondents recruited from CloudResearch in exchange for monetary compensation.[4] Half of the respondents were randomly assigned to the communal brand condition, and the other half were assigned to the exchange brand condition. Within each brand relationship condition (exchange, communal), we manipulated the social activism context (no activism, social activism) between subjects.

Participants completed the main study via an online survey. Respondents were randomly assigned to the communal or the exchange brand condition; they either read the tweet from the communal brand (i.e., Starbucks) or the exchange brand (i.e., Wendy's) from our pretest. Neither tweet directly addressed the social activism context. See Web Appendix C Table W11. Within each brand relationship condition, we varied the social activism context. In the social activism condition, the brand tweet was preceded by a series of six tweets about #MeToo. See Web Appendix C Table W12. In the no activism context, respondents only read the tweet from the brand; it was not preceded by any #MeToo tweets. After reading the tweets, participants completed the same measures of brand attitudes, norm compliance, MeToo movement position, and political ideology as in Study 2. See Web Appendix B Table W3. Finally, participants provided their demographic information.

---

[4] Three respondents were excluded for completing the survey in less than 50 seconds in the control condition and less than 80 seconds in the social activism condition. Average completion time for the remaining respondents was 3 minutes ($SE$ = 9.48 seconds) and 4 minutes ($SE$ = 11.50 seconds), respectively. The reported sample size for Study 3 reflects the exclusion of these respondents.

*Findings*

An ANOVA revealed the predicted Brand × Context interaction on brand attitudes, $F(1, 375) = 7.53$, $p = .01$, $\eta_p^2 = .02$. Planned contrasts within each context condition showed that in the *social activism context*, respondents evaluated the communal brand less favorably ($M = 4.37$, $SE = 0.14$) than the exchange brand ($M = 5.07$, $SE = 0.15$; $t(375) = 3.43$, $p < .01$, $d = .50$). Yet, in the *no activism context*, attitudes did not differ for the communal brand ($M = 5.05$, $SE = 0.15$) versus the exchange brand ($M = 4.95$, $SE = 0.15$; $t(375) = -0.47$, $p = .64$, $d = .07$). Further, the Brand × Context interaction is attributed to less favorable attitudes towards the communal brand in the social activism (vs. no activism) context, $t(375) = 3.28$, $p < .01$, $d = .48$. There was no difference in attitudes for the exchange brand by context, $t(375) = -0.60$, $p = .55$, $d = .09$. See Figure 3.

Figure 3 The Effect of Communal vs. Exchange Brands and Social Activism Context on Brand Attitudes in Study 3.

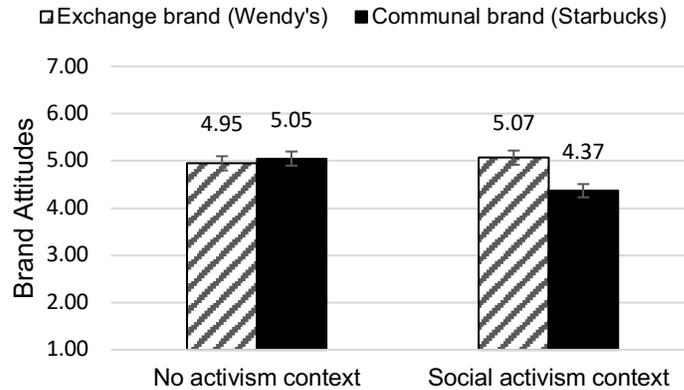

The analysis also showed a significant main effect of brand, with the communal brand ($M = 4.71$, $SE = 0.10$) evaluated less favorably than the exchange brand ($M = 5.01$, $SE = 0.10$), $F(1, 375) = 4.33$, $p = .04$, $\eta_p^2 = .01$. The main effect of social activism context was not significant, $F(1, 375) = 3.58$, $p = .06$, $\eta_p^2 < .01$. Overall, these findings support H1.

*Robustness checks: respondent gender, #MeToo position, and political ideology*

As in Study 2, we conducted additional analyses to examine the role of respondent gender, issue position, and partisanship on our results. We conducted an ANOVA to examine the interactive effect of brand relationship, social activism context, and respondent gender on brand attitudes. The Brand x Context x Gender interaction was not significant, $F(1, 369) = 0.07$, $p = .79$, $\eta_p^2 < .01$. In addition, the Brand × Context interaction on brand attitudes remained significant with the inclusion of #MeToo position as a covariate, $F(1, 374) = 7.48$, $p = .01$, $\eta_p^2 = .02$, and political ideology as a covariate, $F(1, 374) = 7.52$, $p = .01$, $\eta_p^2 = .02$.

*Mediation analysis*

We contend that our effects arise because the social activism context heightens the salience of brand relationship norms. For communal brands, failing to respond to such online conversations is perceived as a violation of these salient norms, whereas no such violation is perceived for exchange brands; therefore, consumers evaluate communal brands less favorably than exchange brands in social activism contexts. Yet, when there is no activism, relationship norms are less salient. This decreases the mediating role of norm compliance on consumers' evaluations of communal and exchange brands. We therefore expected norm compliance to mediate the effect of brand relationship on brand attitudes only in the context of social activism (H2). Consistent with this theorizing, we conducted a mediation analysis with the social activism context conditions (excluding the no activism conditions) using PROCESS, with 5000 bootstrap samples (Model 4; Hayes 2017). See Web Appendix B Table W4 for norm compliance measure means. As predicted, the indirect path of brand relationship (0 = *exchange brand*, 1 = *communal brand*) on brand attitudes through norm compliance for the social activism context conditions was significant ($b = -0.41$, $SE = 0.15$, $95\%$ $CI = [-0.72, -0.12]$). See Figure 4.

Figure 4: Norm Compliance of Response Strategy Mediates the Effect of Communal vs. Exchange Brands and Social Activism Context on Brand Attitudes in Study 3.

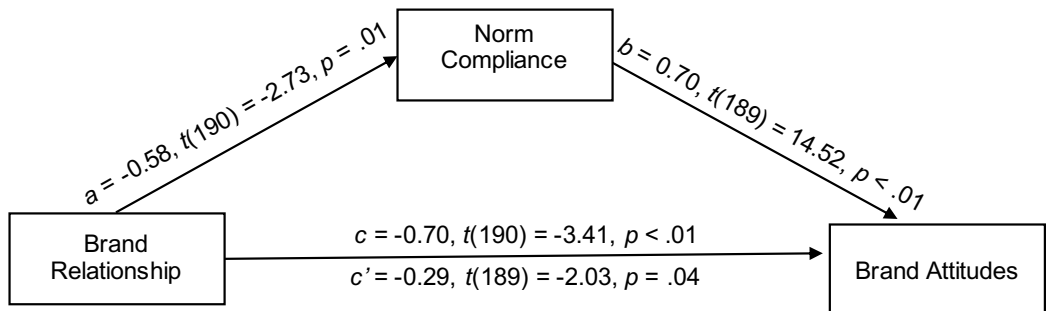

*Notes.* N = 192. Indirect effect: $b = -0.41$, $SE = 0.15$, $95\%$ $CI = [-0.72, -0.12]$.

To further support our theorizing, we conducted an additional mediation analysis (Model 8; Hayes 2017) with brand relationship (0 = *exchange brand*, 1 = *communal brand*), the social activism context (0 = *no activism*, 1 = *social activism*), and their interaction as predictors. In support of our theorizing, the indirect path of brand relationship on brand attitudes through norm compliance was significant for the social activism context ($b = -0.38$, $SE = 0.14$, $95\%$ $CI = [-0.66, -0.10]$), but not for the no activism context ($b = -0.13$, $SE = 0.12$, $95\%$ $CI = [-0.37, 0.11]$).

Together, these findings support our prediction that norm compliance mediates the effect of brand relationship on individuals' brand evaluations in social activism contexts (H2).

**Study 4**

Studies 1-3 focused on one social activism event–i.e., the #MeToo movement. #MeToo relates to gender discrimination, an issue which is generally considered less politically polarizing than other social activism issues–e.g., racial discrimination (Pew Research 2019). Hence, it is possible that our observed effects only arise for social activism events that are not very politically divisive. We therefore conducted Study 4 to ensure that our effects replicate across both low- and high-polarizing events. We selected #MeToo as our low political polarization event, consistent with previous studies, and #BLM as our high polarization social activism event due to its focus on racial issues (Pew Research 2019). Investigating these particular activism events also allows us to better understand our documented effects (i.e., #MeToo) relative to prior work that has investigated brand responses in the context of the #BLM movement (Thomas and Chintagunta 2022; Wang et al. 2022).

*Method*

*Stimulus materials pretest*

Prior to the main study, we ran a pretest using the same respondent sample as the main study (CloudResearch's Connect platform) to confirm that gender issues (#MeToo), are indeed perceived as less politically polarizing than racial issues (#BLM), consistent with prior work (Pew Research 2019). In the pretest ($N = 212$, 49.1% female, 50.5% male, 0.5% prefer not to say, $M_{Age}$ = 39.93, $SD$ = 12.34, Age range = 19-80 years),[5] respondents completed measures of their position on the #MeToo and #BLM movements.

Higher numbers on these measures indicated greater support for these movements (i.e., a more liberal political position). They also completed measures that assessed their stances on both gender discrimination and racial discrimination issues (Pew Research 2019; Westfall et al. 2015). For each of these measures, higher (lower) numbers were indicative of a more conservative (liberal) political position on these issues. These questions were included to ensure that perceptions of the social activism events mirrored those for the issues (i.e., gender and racial discrimination) that they represented. Finally, respondents reported their political beliefs (1 = *extremely liberal*, 4 = *moderate*, 7 = *extremely conservative*) and provided their demographic information. See Web Appendix B Table W6.

Inspired by the past research on political issue polarization (Castle et al. 2020), we ran a correlation analysis to assess the relationship between political beliefs and respondents' position on #MeToo and #BLM. This analysis revealed significant, negative correlations for both the #MeToo ($r$ = -.55) and #BLM ($r$ = -.69) movements. However, the magnitude of correlation was lower for the gender versus the race issue, $z$ = 2.22, $p$ = 0.03, indicating a less political polarization. The correlations between political beliefs and stance on each issue (gender discrimination, racial discrimination) showed a similar pattern of results ($r_{Gender}$ = .58, vs. $r_{Race}$ = 0.69), $z$ = 1.86, $p$ = .07. We also conducted an ANOVA with political beliefs as a between-subjects variable, social activism event (#MeToo vs. #BLM) as a within-subjects variable, and movement position as our dependent measure. In line with prior work (Pew Research 2019), we grouped respondents into 3 categories based on their political beliefs (Liberal/lean liberal: scale points 1-3, Moderate/middle-of-the-road: scale point 4, and Conservative/lean conservative: scale points 5-7). This analysis revealed a significant Political belief × Event position interaction, $F(2, 209) = 13.44$, $p < .01$, $\eta_p^2 = .11$; liberal and conservative respondents exhibited a greater disparity in their stances on #BLM ($M_{Liberal} = 5.75$, $SE = 0.17$ vs. $M_{Conservative} = 2.39$, $SE = 0.21$), $t(171) = 13.20$, $p < .01$, $d = 2.06$, than #MeToo ($M_{Liberal} = 5.66$, $SE = 0.17$ vs. $M_{Conservative} = 3.55$, $SE = 0.21$), $t(171) = 7.49$, $p < .01$, $d = 1.23$. See Web Appendix B Table W7. Combined, these findings confirm that the #BLM (vs. #MeToo) movement is more politically polarizing.

---

[5] Two respondents in the Study 4 pretest were excluded from the analysis for completing the survey in more than 14.5 minutes. Average completion time for remaining respondents was 2.1 minutes ($SE$ = 6.48 seconds). The reported sample size for the pretest reflects the exclusion of these respondents.

*Participants, design and procedure*

Participants ($N = 623$, 49.6% female, 50.1% male, 0.3% prefer not to say, $M_{Age} = 39.22$, $SD = 11.95$, Age range = 18-85 years) were U.S.-based respondents recruited from CloudResearch's Connect platform.[6] We manipulated brand relationship (communal, exchange) and the political polarization of the social activism event (low polarization, high polarization) between subjects. In addition, we included control conditions in which there was no social activism. This was done to mirror the no activism conditions in Study 3 and, more importantly, to offer additional insight into the relative impact of social activism events with low versus high political polarization. Thus, a total of six experimental conditions were used.

Participants completed the main study via an online survey. As in Study 3, respondents were randomly assigned to read a tweet for the communal (i.e., Starbucks) or exchange (i.e., Wendy's) brand. Neither tweet directly addressed the social activism context; these were the same tweets as those used in Study 3. Respondents in the control conditions simply read the brand tweet and completed the dependent measures. In the event polarization conditions (i.e., #MeToo, #BLM), the brand tweet was preceded by a series of tweets that referenced the event. See Web Appendix C Tables W12 and W13. Participants in these conditions read that they were scrolling on Twitter and came across a trending conversation. They either read that the trending conversation pertained to #MeToo (low political polarization event) or #BLM (high political polarization event, after which they read six tweets about this event. After reading the event tweets and then the brand tweet, they completed the dependent measures. All participants completed the same brand attitudes and norm compliance measures from previous studies. Next, they completed measures of their position on the issue and political ideology (liberal vs. conservative; Smith et al. 2018) to account for any individual differences related to the two issues that we investigated (i.e., #MeToo, #BLM). Finally, participants provided their demographic information.

---

[6] Seven respondents were excluded from the analysis for completing the survey in less than 60 seconds. Average completion time for remaining respondents was 3.2 minutes ($SE = 5.07$ seconds). The reported sample size for Study 4 reflects the exclusion of these respondents.

*Findings*

We present the analyses for the two event polarization conditions first. We then present the relevant planned contrasts for the control (i.e., no activism context) conditions.

An ANOVA revealed a main effect of brand relationship; the communal brand ($M = 4.49$, $SE = 0.09$) was evaluated less favorably than the exchange brand ($M = 4.97$, $SE = 0.09$), $F(1, 415) = 12.94$, $p < .01$, $\eta_p^2 = .03$. This finding parallels the pattern of results from the social activism context conditions in Study 3. The analysis also showed a significant main effect of event polarization, such that respondents' brand attitudes were less favorable with the low polarization event (#MeToo; $M = 4.59$, $SE = 0.10$) than with high polarization event (#BLM; $M = 4.87$, $SE = 0.09$), $F(1, 415) = 4.67$, $p = .03$, $\eta_p^2 = .01$. The analysis did not show a significant Brand × Event polarization interaction on brand attitudes, $F(1, 415) = 0.61$, $p = .43$, $\eta_p^2 < .01$.

This finding supports our theorizing that the predicted effect of brand relationship (communal vs. exchange) persists across social activism events that differ in their political polarization.

We conducted planned contrasts within each event polarization condition to further investigate their relative effects. With the *low polarization event* (#MeToo), respondents evaluated the communal brand less favorably ($M = 4.40$, $SE = 0.14$) than the exchange brand ($M = 4.77$, $SE = 0.14$), $t(415) = 1.94$, $p = .05$, $d = 0.28$. The *high polarization event* (#BLM) showed the same pattern of results; respondents evaluated the communal brand less favorably ($M = 4.58$, $SE = 0.13$) than the exchange brand ($M = 5.17$, $SE = 0.13$), $t(415) = 3.19$, $p < .01$, $d = 0.43$. These results showed a similar pattern for the low versus high polarization events, although the effect size was larger for the high political polarization event (#BLM). Further, the difference in the effect size is due to a divergence in evaluations of the exchange brand (Wendy's), and not the communal brand (Starbucks), in the context of #BLM versus #MeToo. Respondents evaluated the exchange brand less favorably in the context of the low (vs. high) polarization event, $t(415) = 2.07$, $p = .04$, $d = 0.29$. Yet, evaluations of the communal brand did not differ by event polarization, $t(415) = 0.98$, $p = .33$, $d = 0.14$. See Figure 5. Overall, these findings provide further evidence for H1; the predicted effect of brand relationship (communal vs. exchange) persists across various social activism events, regardless of their political polarization.

Figure 5: The Effect of Communal vs. Exchange Brands and Political Polarization of Social Activism Events on Brand Attitudes in Study 4.

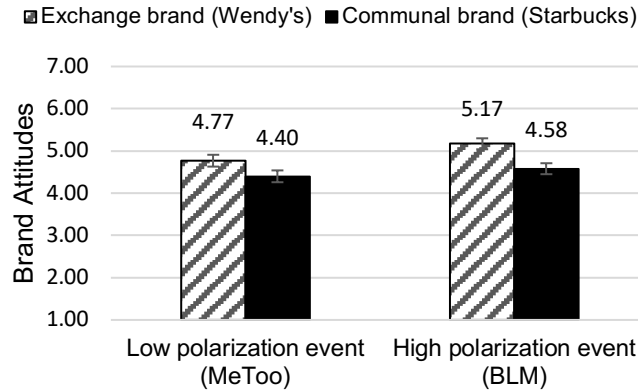

*Robustness checks: respondent gender, respondent ethnicity, issue position, and political ideology*

As in previous studies, we conducted additional analyses to examine the role of individual differences for respondents (i.e., gender, ethnicity, issue position, and political ideology) on our effects. See Web Appendix B Table W9 for details on the respondent sample. We conducted an ANOVA to examine the effect of the Brand x Event interaction on brand attitudes with all four variables included as controls. The main effect of brand relationship remained significant with the inclusion of these covariates, $F(1, 411) = 13.97$, $p < .01$, $\eta_p^2 = .03$. In addition, we conducted four separate ANOVAs to examine the interaction between brand relationship and (1) respondent gender, (2) respondent ethnicity, (3) issue position, and (4) political ideology on brand attitudes. We did not include social activism event as a factor in these analyses since #MeToo and #BLM yielded a similar pattern of results. The effects of Brand x Gender, $F(1, 414) = 0.62$, $p = .43$, $\eta_p^2 < .01$, Brand x Ethnicity, $F(3, 410) = 0.33$, $p = .80$, $\eta_p^2 < .01$, Brand x Issue position, $F(1, 415) = 2.67$, $p = .10$, $\eta_p^2 = .01$, and Brand x Political ideology, $F(1, 415) = 0.07$, $p = .93$, $\eta_p^2 < .01$, were not significant. These results suggest the effects are not attributable to individual differences in demographics, issue opinions, or political stance.

*Mediation analysis*

Consistent with Studies 2 and 3, we expected norm compliance to mediate our effects. We expected that event polarization would not impact the relationship between brand relationship and norm compliance; rather, any effect would be limited to the relationship between brand relationship and brand attitudes. We conducted the mediation analyses using PROCESS, with 5000 bootstrap samples (Model 5; Hayes 2017). In our analysis, brand relationship (0 = *exchange brand*, 1 = *communal brand*) was the predictor, the social activism event polarization (0 = *low event polarization*, 1 = *high event polarization*) was the moderator, and brand attitudes was the dependent variable. See Web Appendix B Table W4 for norm compliance means. As predicted, the indirect path of brand relationship on brand attitudes through norm compliance was significant ($b = -0.37$, $SE = 0.10$, $95\%$ $CI = [-0.57, 0.-0.17]$). See Figure 6. These findings offer additional support for the mediating role of norm compliance in our observed effects (H2).

Figure 6: Norm Compliance of Response Strategy Mediates the Effect of Communal vs. Exchange Brands on Brand Attitudes in Study 4.

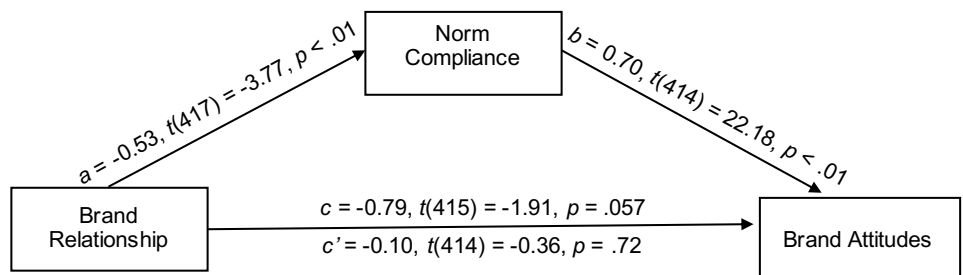

*Notes.* N = 419. Indirect effect: b = -0.37, *SE* = 0.10, *95% CI* = [-0.57, 0.-0.17].

*Comparisons to control conditions*

Our findings show that the divergence in evaluations for communal versus exchange brands in the wake of social activism persists for events with low and high political polarization.

We expected that the pattern occurs because communal brands are penalized more than exchange brands for not directly respond to social activism, regardless of the political polarization of the event. Replicating Study 3, in the control conditions in which there is no social activism, there were no significant differences in brand attitudes between the communal ($M = 5.03$, $SE = 0.13$) and exchange brand ($M = 4.96$, $SE = 0.14$), $t(617) = -0.36$, $p = .72$, $d = .05$. Yet, as predicted, for the exchange brand (Wendy's), evaluations did not differ significantly from the control for both the low polarization event, $t(617) = 0.98$, $p = .33$, $d = .14$, and the high polarization event, $t(617) = -1.07$, $p = .29$, $d = .15$. However, evaluations of the communal brand (Starbucks) decreased relative to the control for both the low polarization event (#MeToo), $t(617) = 3.30$, $p < .01$, $d = .46$, and the high polarization event (#BLM), $t(617) = 2.42$, $p = .02$, $d = .33$. In general, consumers evaluate communal brands less favorably when they do not engage in social activism; these findings appear to be robust across various types of social activism events.

## Study 5

The objective of Study 5 was to explore the effectiveness of brand responses to social activism for communal versus exchange brands (H3). We predict that a communal brand will be evaluated less favorably than an exchange brand when using a low empathy response in the context of social activism, replicating the findings from Study 3 social activism conditions in which the brand does not directly respond to the #MeToo conversation. Yet, we expect that the difference in evaluations of communal versus exchange brands will be attenuated when using a high empathy response.

### *Method*

### *Stimulus materials pretest*

In Study 5, respondents were asked to read a tweet from a communal or exchange brand. However, we also varied the brand's response to social activism, such that the brand responded to the #MeToo conversation with either a low empathy or high empathy tweet. We used the same brands as those used in Studies 3 and 4 (i.e., Wendy's and Starbucks). Yet, prior to the main study, we wanted to confirm that our brand responses affected perceived empathy as intended and not other constructs. We conducted a pretest using the same respondent sample as the main study (CloudResearch).

In the pretest ($N = 119$, 47.1% female, $M_{Age} = 38.56$, $SD = 11.12$, Age range = 20-70 years), we sought to verify that the brand response (i.e., low vs. high empathy) was affecting perceived brand empathy as intended and *not* inadvertently affecting believability of the response. All participants read the six #MeToo tweets from the social activism conditions in Study 3, followed by one brand tweet. Participants were randomly assigned to read one of two versions of the brand tweet. In the low empathy version, the brand simply referred individuals to a link to learn more about the brand's #MeToo statement (i.e., acknowledging others' issues). In the high empathy version, this statement was preceded by another statement about the brand's efforts to change practices and support cultural reforms (i.e., acting to address the issue). This way of designing control and treatment text responses is consistent with prior research (Lei, Yin, and Zhang 2021). We did not vary the brand in this pretest; we only referred to the brand as XYZ since we wanted to understand the perceptions of the different responses in isolation (i.e., without the impact of a priori brand relationships). See Web Appendix C Table W14.

After reading the tweets, participants completed measures of perceived empathy (Batson, Fultz, and Schoenrade 1987; Coke, Batson, and McDavis 1978) and response believability (Cowen and Montgomery 2020). We include the believability measure because the existing literature on brand engagement in sociopolitical issues has documented the importance of authenticity in engagement (Moorman 2018; Vredenburg et al. 2020). See Web Appendix B Table W8 for details on pretest measures. An ANOVA with response type as the factor showed a significant main effect on perceived empathy, $F(1, 117) = 6.55$, $p = .01$, $\eta_p^2 = .05$. Participants rated the high empathy response as more empathetic ($M = 5.26$, $SE = 0.17$) than the low empathy response ($M = 4.64$, $SE = 0.17$). Yet, there were no differences in response believability, $F(1, 117) = 0.86$, $p = .36$, $\eta_p^2 = .01$. These results increase our confidence that any moderating effects of response type are indeed attributable to our proposed mechanism (high vs. low empathy response) and not the differences in perceived believability.

*Participants, design and procedure*

Study 5 ($N = 706$, 51.6% female, $M_{Age} = 39.30$, $SD = 12.12$, Age range = 21-78 years) used U.S.-based respondents recruited from CloudResearch.[7] We manipulated brand relationship (exchange, communal) and response type (low empathy response, high empathy response) between subjects. In addition, we included control conditions in which the brand did not directly respond to the #MeToo conversation. This was done to mirror the social activism conditions in Studies 3-4 and, more importantly, to offer additional insight into the impact of a low versus high empathy response relative to when brands do not respond directly to social activism. Thus, a total of six experimental conditions were used.

Study 5 was completed via an online survey. Participants were randomly assigned to read the control, low empathy, or high empathy response tweet from either a communal (i.e., Starbucks) or exchange (i.e., Wendy's) brand. The low and high empathy tweets were the same as those used in the Study 5 pretest, except that XYZ was replaced by brand names. The control tweets were the same as those used in Study 3. After reading the six #MeToo tweets and the randomly-assigned brand tweet, participants completed the same brand attitudes and norm compliance measures as in Studies 2-4 to facilitate comparisons of these measures to the response conditions. In addition, they reported their purchase intent (Cowen and Montgomery 2020; Park et al. 2010) and WOM (Chen and Berger 2016; Montgomery and Cowen 2020). They also completed a series of questions to assess their likelihood of engaging in behaviors on Twitter (He, Rui, and Whinston 2018) after reading the brand's response to the #MeToo conversation (i.e., *retweet post*, *comment on post*, *like post*, *follow brand*, and *click on link in post*). See Web

---

[7] Four respondents in Study 5 were excluded from the analysis for completing the survey in less than 80 seconds. Average completion time for remaining respondents was 4.2 minutes ($SE = 7.28$ seconds). The reported sample size for Study 5 reflects the exclusion of these respondents.

Appendix B Table W8. We included these additional measures in Study 5 to better understand other consequential behaviors studied in prior work (Wang et al. 2022) that may be impacted by a brand responding to a social activism conversation on Twitter. Finally, all participants provided their demographic information.

*Findings*

We present the analyses for the two social activism response conditions first. We then present the relevant planned contrasts for the control (i.e., no response) conditions.

An ANOVA revealed a main effect of brand relationship; the communal brand ($M$ = 4.99, $SE$ = 0.08) was evaluated less favorably than the exchange brand ($M$ = 5.26, $SE$ = 0.08), $F(1, 476)$ = 5.12, $p$ = .02, $\eta_p^2$ = .01. This finding parallels the pattern of results from the social activism context conditions in Study 3 and 4. The analysis also showed a significant main effect of response type, such that respondents' brand attitudes were less favorable when a low (vs. high) empathy response was used ($M_{\text{Low empathy}}$ = 4.80, $SE$ = 0.08 vs. $M_{\text{High empathy}}$ = 5.45, $SE$ = 0.08), $F(1, 476)$ = 30.21, $p$ < .01, $\eta_p^2$ = .06. Importantly, the analysis revealed the predicted Brand × Response type interaction on brand attitudes, $F(1, 476)$ = 6.04, $p$ = .01, $\eta_p^2$ = .01. Planned contrasts within each response condition supported our theorizing. When a *low empathy response* was used, respondents evaluated the communal brand less favorably ($M$ = 4.52, $SE$ = 0.12) than the exchange brand ($M$ = 5.08, $SE$ = 0.12), $t(476)$ = 3.34, $p$ < .01, $d$ = 0.43. Yet, when the *high empathy response* was used, respondents rated the two brands similarly ($M_{\text{Exchange brand}}$ = 5.43, $SE$ = 0.12 vs. $M_{\text{Communal brand}}$ = 5.46, $SE$ = 0.12), $t(476)$ = -0.14, $p$ = .89, $d$ = 0.02. Further, the attenuation in the difference of attitudes towards the communal versus exchange brand is due to a larger increase in brand attitude for the high (vs. low) empathy response condition for the communal brand context, $b$ = .94, $SE$

= .17, $t(476) = -5.60$, $p < .01$, $d = .73$, than for the exchange brand, $b = .36$, $SE = .17$, $t(476) = -2.16$, $p = .03$, $d = .28$. See Figure 7. Overall, these findings support H3.

Figure 7: The Effect of Communal vs. Exchange Brands and Social Activism Response Type on Brand Attitudes in Study 5.

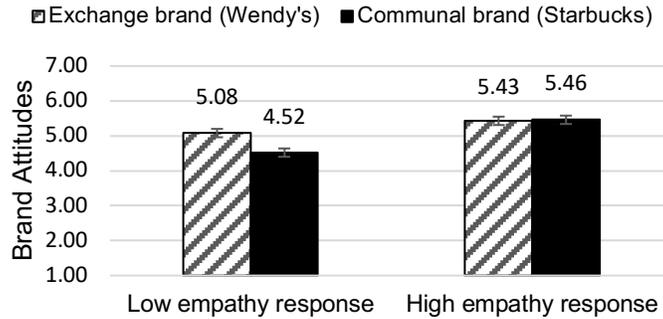

Table 2: Other Dependent Measure Means by Condition in Study 5.

| Conditions | Purchase intent | WOM | Retweet post | Comment on post | Like post | Follow brand | Click link |
|---|---|---|---|---|---|---|---|
| Low empathy response | | | | | | | |
|   Exchange brand | 4.90 (.16) | 4.93 (.15) | 2.24 (.16) | 2.20 (.16) | 3.10 (.20) | 2.81 (.18) | 4.10 (.20) |
|   Communal brand | 3.97 (.16) | 3.98 (.15) | 1.78 (.16) | 2.04 (.16) | 2.50 (.20) | 2.14 (.18) | 3.46 (.20) |
| High empathy response | | | | | | | |
|   Exchange brand | 5.04 (.16) | 5.12 (.15) | 2.72 (.16) | 2.76 (.16) | 3.88 (.20) | 3.14 (.18) | 3.64 (.20) |
|   Communal brand | 4.81 (.16) | 4.92 (.15) | 2.65 (.16) | 2.57 (.17) | 3.86 (.20) | 3.17 (.18) | 3.51 (.20) |

*Notes.* Standard errors are denoted in parentheses. One respondent did not complete the 'Follow brand' measure.

The analysis of the purchase intent, WOM, and social media engagement measures revealed a similar pattern of results. See Table 2. Interestingly, after reading a low empathy response, respondents expressed a lower level of social media engagement (i.e., retweet, like, and follow) for the communal (vs. exchange) brand. However, this effect was attenuated in the high empathy response condition. These findings suggest that a high empathy response is more effective and engaging for a communal brand, further confirming our theorizing.

*Robustness checks: respondent gender, #MeToo position, and political ideology*

As with the previous studies, we conducted additional analyses to examine the role of respondent gender, issue position, and political ideology on our results. We conducted an ANOVA to examine the interactive effect of brand relationship, response type, and respondent gender on brand attitudes. The Brand x Response x Gender interaction was not significant, $F(1, 472) = 0.11$, $p = .74$, $\eta_p^2 < .01$. In addition, the Brand × Response interaction remained significant with the inclusion of #MeToo position as a covariate, $F(1, 475) = 5.45$, $p = .02$, $\eta_p^2 = .01$, and political ideology as a covariate, $F(1, 475) = 5.99$, $p = .02$, $\eta_p^2 = .01$. These results further confirm that our hypothesized effects were not driven by gender, issue stance, or partisanship.

*Mediation analysis*

We expected norm compliance to mediate the effects of brand relationship and response type on brand evaluations (H2). In our analysis, brand relationship (0 = *exchange brand*, 1 = *communal brand*), response type (0 = *low empathy response*, 1 = *high empathy response*), and their interaction were predictors, and brand attitudes was the dependent variable.[8] See Web Appendix B Table W4 for norm compliance means. We conducted a mediation analysis using PROCESS, with 5000 bootstrap samples (Model 8; Hayes 2017). The index of moderated mediation ($b = 0.35$, $SE = 0.17$, $95\% CI = [0.02, 0.69]$) was significant. Also, as predicted, the indirect path of social activism on brand attitudes through norm compliance for the *low empathy response* was negative and significant ($b = -0.29$, $SE = 0.13$, $95\% CI = [-0.54, -0.05]$), mirroring the results from the social activism context in Study 3. Yet, the indirect path for *high empathy response* was not significant (*b* = 0.05, *SE* = 0.11, *95% CI* = [-0.18, 0.28]), similar to the context in Study 3 where there was no social activism. These findings further support the mediating role of norm compliance in our observed effects. See Figure 8.

---

[8] The mediation models with purchase intent (index of moderated mediation: $b = 0.38$, $SE = 0.19$, $95\% CI = [0.02, 0.76]$, and WOM (index of moderated mediation: $b = 0.38$, $SE = 0.19$, $95\% CI = [0.03, 0.75]$).

Figure 8: Norm Compliance of Response Strategy Mediates the Effect of Communal vs. Exchange Brands and Social Activism Response Type on Brand Attitudes in Study 5.

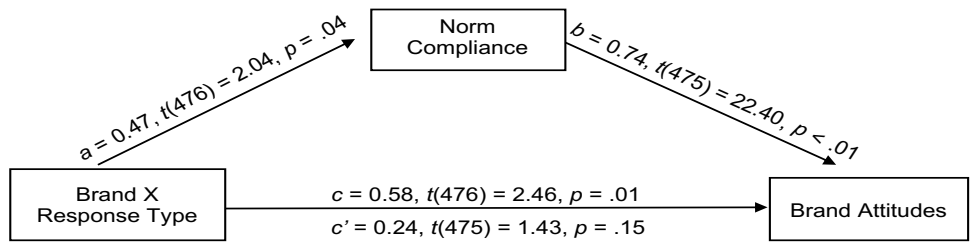

*Notes.* N = 480. Index of moderated mediation: b = 0.35, SE = 0.17, 95% CI = [0.02, 0.69].
*Comparisons to control conditions*

      Our findings show that evaluations for communal versus exchange brands differ when a low empathy response is utilized in the wake of social activism, and this divergence in evaluations is attenuated with a high empathy response. We expected that the pattern occurs because, for a communal brand, a low empathy response does not alter evaluations relative to when the brand does not directly respond to social activism, whereas a high empathy response increases evaluations. Further, the increase in evaluations with a high empathy response should be greater for a communal brand than for an exchange brand, resulting in the attenuation of the difference in brand evaluations. Replicating Study 3, in the control conditions in which the brand did not directly respond to the #MeToo conversation, the communal brand was evaluated less favorably (*M* = 4.38, *SE* = .12) than the exchange brand (*M* = 4.72, *SE* = .12), *t*(700) = 1.99, *p* = .05, *d* = .27. As expected, for the communal brand, we found no significant differences in brand attitudes between the low empathy response and the control/no response condition, *t*(700) = 0.81, *p* = .42, *d* = .11, but the high empathy response increased evaluations relative to the

control, $t(700) = 6.30$, $p < .01$, $d = .57$. However, for the exchange brand, brand evaluations increased relative to the control for both the low empathy response $t(700) = 2.09$, $p = .04$, $d = .27$, and the high empathy response $t(700) = 4.24$, $p < .01$, $d = .55$, although the high empathy response yielded a larger increase in evaluations ($b_{\text{High empathy}} = 0.71$ vs. $b_{\text{Low empathy}} = 0.35$). In addition, the increase in evaluations in the high empathy response conditions was greater for the communal brand ($b = 1.08$) than the exchange brand ($b = 0.71$), offering additional support for our theorizing. Overall, our findings suggest that differences in consumers' evaluations of communal versus exchange brands in a social activism context can be attenuated by a high empathy response but not a low empathy response (H3). This is because a high empathy response is perceived to comply with brand relationship norms for communal brands (H2).

## Discussion

Across five studies, we clarify how brand response strategy (i.e., whether and how to respond) to social activism impacts consumers' evaluations of communal versus exchange brands. When brands do not respond to social activism, consumers evaluate communal (vs. exchange) brands less favorably (H1; Studies 1-3). This effect persists across social activism events that differ in their political polarization (H1; Study 4). This divergence in evaluations is attenuated when brands utilize a high empathy response to engage in social activism, but not a low empathy response (H3; Study 5). We attribute these findings to differences in the extent to which response strategies of communal versus exchange brands are perceived to comply with relationship norms during social activism (H2; Study 2-4). These effects consistently replicate across studies and various dependent measures, offering robust support for our predictions and proposed theoretical mechanism. We also rule out several competing explanations for the findings, including demographics, issue stance, and political ideology.

*Theoretical Implications*

Our research contributes to the literature on brand engagement in social activism (e.g., Bhagwat et al. 2020; Moorman 2020; Sarkar and Kotler 2018; Vredenburg et al. 2020; Wang et al. 2022). By revealing the underlying mechanism of these effects (i.e., compliance with brand relationship norms), our work may explain the mixed findings on whether brand engagement in social activism is beneficial or harmful for brands (Bhagwat et al. 2020; Chatterji and Toffel 2019; Vredenburg et al. 2020; Wang et al. 2022). For example, the negative effects of brand response to #BLM on social media engagement observed in Wang et al. (2022) could be attributed to brand responses (e.g., low empathy response) and/or brand composition (e.g., the sample is comprised of many communal brands).

In addition, prior work on brand engagement in social issues has primarily focused on consumers' reactions to partisanship issues, often examining consumers' brand evaluations as a function of their political affiliation (Hydock, Paharia, and Blair 2020; Mukherjee and Althuizen 2020; Wang et al. 2022). Our findings suggest that the consumer-brand relationship is more impactful on consumers' brand evaluations than political affiliation. We do not observe differences in political ideology in our studies, and the brand relationship effects persist across events with varying levels of political polarization.

Furthermore, we empirically document the benefits to brands of engaging in social activism (i.e., more favorable brand attitudes, higher purchase intent, better WOM, and more social media engagement). In doing so, we complement prior work focused on the societal (vs. brand) benefits of brand involvement in these issues (Den Hond and De Bakker 2007; Du, Bhattacharya, and Sen 2011; Lin 2018).

Our research also extends work on brand relationship norms. Work in this domain has primarily focused on how actions internally controlled by brands (e.g., customer service issues and product failures) violate or conform to salient relationship norms (Aggarwal 2004, Aggarwal and Larrick 2012, Aggarwal and Law 2005). To our knowledge, our research is the first to show that actions *not* internally controlled by brands (i.e., social activism events) can still affect perceptions of the brand's norm compliance and, consequently, consumers' evaluations of brands. Thus, we expand the boundary condition of brand relationship norms theory.

Finally, our work extends the crisis communication literature (Bundy et al. 2017; Bundy and Pfarrer 2015; Coombs and Holladay 2002, 2010). Research in this domain often focuses on responses to crisis conditions that are related to an organization's core businesses (e.g., product failure, product recall). Our findings illustrate additional crisis conditions—i.e., social activism that is not directly under a brand's control—for which consumers still expect brands to respond and be accountable. This perceived accountability is due to a priori consumer-brand relationship and corresponding expectations. With this, we suggest an effective response strategy–i.e., high empathy response–to match this perceived degree of culpability for communal brands, thus adding a new crisis response strategy to this literature (Coombs and Holladay 2002).

*Practical Implications*

Our work also offers additional insight to marketing and brand managers on how to build and sustain communal relationships with consumers. Many communal brands have already engaged in societal issues because it is a crucial way to maintain their communal relationships with consumers (Hale 2021). However, these brands must often use a process of trial and error to appropriately tailor their response strategies to various social activism movements. By showing how consumer evaluations of brands are impacted by social activism, we are able to elucidate the conditions under which brands should respond and how to craft an appropriate response. This knowledge will ultimately help communal brands save valuable time and resources. On the other hand, many exchange brands also strive to build more communal relationships with consumers due to the tremendous benefits of such relationships (Aaker, 1997; Aaker et al., 2004; Ahluwalia et al., 2000; Fournier, 1998; Keller, 1993). We provide a new pathway for them to support these efforts—i.e., engaging in social activism. In fact, in Study 5, we find that responding to social activism, compared to not responding at all, may also yield benefits for exchange brands, albeit to a lesser extent than for communal brands.

Additionally, our research assists social advocates in their efforts to garner more brand support for their issues. Brand involvement in social activism has a multitude of societal benefits (Lin 2018). However, brands are reluctant to take a public stance on social issues due to the high level of uncertainty associated with such engagement. Our findings trigger more brand involvement by documenting the concrete benefits of such involvement (or the perils of lack of involvement) to brands themselves—i.e., consumers evaluate brands more favorably if they do (vs. do not) get involved. We also alleviate brands' concerns in engaging in societal issues by prescribing concrete response strategies to help brands engage more effectively. Through this two-pronged approach, we help social activists solicit additional brand support to deepen and broaden social changes to fight injustice.

**Limitations, Future Directions, and Conclusion**

Future research can build on our research in several ways. First, in addition to investigating how brand response strategy for social activism can violate or conform to salient relationship norms, future inquiries can also examine how a mild deviation from a priori relationships and expectations can "delight" or pleasantly surprise consumers, leading to favorable brand evaluations. Vredenburg et al. (2020) propose (although have not empirically validated) that when a brand's engagement in sociopolitical issues diverges from norms and expectations in a bounded, not extreme way, such engagement can still yield positive responses – i.e., consumer delight. Our Study 5 also finds that for an exchange brand, an empathetic response to social activism leads to a significant increase in brand evaluations compared with a no-response condition, but the magnitude of such an increase is significantly smaller than for a communal brand. Future research should investigate the nuances and boundary conditions of this effect and add a third dimension to our norm compliance mechanism.

Second, future research could explore other outcomes of brand response strategy during social activism. Brand attitudes was our focal measure, but Study 5 also investigates the effects on purchase intent, WOM, and social media engagement. All measures were selected to mirror prior work on consumer reactions to brands (e.g., Aggarwal 2014; Wang et al. 2022). Yet, we anticipate that our effects will extend to other outcomes that rely on consumer support, given the link between consumers' attitudes and behaviors (Ajzen 1991; Ajzen et al. 2018; Glasman and Albarracín 2006). Thus, future work can examine the robustness of our effects on firm-level outcomes, such as product sales (Mu et al. 2022) and firm value (Chung et al. 2020).

Third, additional research could delve further into the differences in how consumers react to brand responses in the wake of low versus high politically-polarizing events. While our Study 4 reveals a consistent trend across both the #MeToo and #BLM movements, the effect size was significantly larger for the latter. Our findings show that this difference in effect size is due to a divergence in evaluations of the exchange brand, not the communal brand – i.e., consumers evaluated the exchange brand less favorably in the context of the low (vs. high) polarization event. It could be that the #MeToo movement, addressing gender inequality, garners broader support, leading consumers to anticipate responses from both communal and exchange brands, albeit a higher expectation for communal brands. Conversely, it is possible that because #BLM is an issue that divides consumers along party lines, it solicits a tempered consumer expectation of broad brand support. Further exploration in this area could emphasize the significance of brand engagement in social activism, particularly for issues that are less politically polarizing.

Finally, future work should examine the consequences of other aspects of brand responses, such as social media functionalities, print advertisement, news articles, and public announcement, during social activism. For example, prior research has highlighted the dialogic functionality of social media in communications (He, Rui, and Whinston 2018; Rishika et al. 2013). For example, retweeting, tagging, commenting, and multimedia features on Twitter enable organizations to quickly provide a personalized marketing message or service response to consumers (He, Rui, and Whinston 2018). These two-way conversations between consumers and brands might conform more to communal norms by increasing consumer perceptions that brands genuinely care about and support all members of society. Future work might investigate how these social media functionalities impact perceived compliance with relationship norms and, consequently, affect consumers' brand evaluations.

Overall, this research reveals the delicate role of brand relationships in the interplay of consumer responses and brand response strategies in the wake of social activism. Communal brands, despite their robust consumer ties and favorable outcomes, can also face increased backlash when not responding to social activism in line with consumers' expectations. Exchange brands, conversely, can be more intentional with the types of social issues that they address since consumers' have lower expectations that these brands will engage in these online conversations. By elucidating when and how brands should engage in social activism, we hope to garner more brand support for social issues, and, in turn, catalyze societal progress in combatting prejudice and discrimination. Ultimately, our findings underscore the imperative for businesses to balance both instrumental and humanitarian objectives in their operations.